\documentclass[11pt,a4paper]{article}

\pdfoutput=1
\usepackage{jheppub}
\usepackage{amsmath}
\usepackage{mathrsfs}
\usepackage{bm}
\usepackage{epsfig}
\usepackage{amssymb}
\usepackage{graphics}
\usepackage[active]{srcltx}
\usepackage{amsthm}
\usepackage{tcolorbox}
\usepackage{tikz}
\usepackage{tabularx}
\usepackage{algorithm}
\usepackage{algpseudocode}
\usepackage{shuffle}

\newcommand{\RNum}[1]{\uppercase\expandafter{\romannumeral #1\relax}}

\theoremstyle{plain} 
\theoremstyle{plain} 
\theoremstyle{plain} 
\theoremstyle{definition} 
\theoremstyle{definition}

\newcommand{\te}[1]{\text{#1}}

\newcommand{\bea}{\begin{eqnarray}}
\newcommand{\eea}{\end{eqnarray}}
\newcommand{\nn}{\nonumber \\}

\newcounter{protocol}

\def\Tr{\mathop{\rm Tr}}

\def\eref#1{(\ref{#1})}

\title{A new recursion relation for tree-level NLSM amplitudes based on hidden zeros}
\author[a]{Xiaodi Li} \author[b]{Kang Zhou}

\affiliation[a]{Department of Physics, Fudan University, Shanghai 200433, China}

\affiliation[b]{Center for Gravitation and Cosmology, College of Physical Science and Technology, Yangzhou University,\\
No.180, Siwangting Road, Yangzhou, 225009, P.R. China}

\emailAdd{lixiaodi@fudan.edu.cn}
\emailAdd{zhoukang@yzu.edu.cn}

\date{\today}

\abstract{
In this note, we propose a novel BCFW-like recursion relation for tree-level non-linear sigma model (NLSM) amplitudes, which circumvents the computation of boundary terms by exploiting the recently discovered hidden zeros. Using this recursion, we reproduce three remarkable features of tree-level NLSM amplitudes: (i) the Adler zero, (ii) the $\delta$-shift construction, which generates NLSM amplitudes from ${\rm Tr}(\phi^3)$ amplitudes, and (iii) the universal expansion of NLSM amplitudes into bi-adjoint scalar amplitudes. Our results demonstrate that the hidden zeros, combined with standard factorization on physical poles, uniquely determine all tree-level NLSM amplitudes.
}


\begin{document}

\maketitle

\section{Introduction}
\label{sec-intro}

A central idea of the modern $S$-matrix program is to construct scattering amplitudes directly from general physical principles---such as locality and unitarity---without relying on the traditional Lagrangian formulation and Feynman rules. A prominent example is the well-known Britto-Cachazo-Feng-Witten (BCFW) recursion relation \cite{Britto:2004ap,Britto:2005fq}, which employs factorization on physical poles to recursively build higher-point amplitudes from lower-point ones. The BCFW recursion relation has proven to be a powerful tool for computing gluon and graviton amplitudes at tree level. However, when applied to effective field theories (EFTs) such as the non-linear sigma model (NLSM), it encounters a major obstacle: the presence of non-vanishing boundary terms at infinity. For instance, NLSM amplitudes with even external lines always contain contact terms, which do not have any propagator and cannot be factorized. Given the central role of EFTs in describing phenomena such as spontaneous symmetry breaking, it is both natural and important to extend the BCFW recursion to improve its applicability to a broader class of theories, including EFTs.

The presence of boundary terms implies that factorization on physical poles alone is insufficient to fully determine EFT amplitudes, and thus an additional physical principle is required. It is natural to expect that this difficulty can be overcome by invoking such a principle. The first attempt in this direction was made in \cite{Cheung:2015ota}, where the supplementary principle was the existence of special zeros in tree-level EFT amplitudes, known as \emph{Adler zeros}. By appropriately incorporating the Adler zero condition, the boundary terms could be eliminated, enabling a novel BCFW-like recursion relation that was successfully applied to the tree amplitudes of EFTs such as the NLSM, special Galileon, and Born-Infeld theories. However, a drawback of this approach is that it requires the spacetime dimension to satisfy $D \leq 2n-1$, where $2n$ is the total number of external particles---a constraint that conflicts with the expected $D$-independence of purely bosonic EFTs.
More recently, the authors in \cite{Cachazo_2022} proposed another recursion relation for tree-level NLSM amplitudes, where the additional ingredient was a novel factorization property known as \emph{smooth splitting}: under certain conditions, each tree-level NLSM amplitude factorizes into three amputated currents. The recursion relation does not contain the boundary terms by exploiting the smooth splitting and remains insensitive to the spacetime dimension, since the Mandelstam variables, rather than the momenta of external particles, are directly shifted in \cite{Cachazo_2022}. Nevertheless, this recursion relation involves amputated currents, whose evaluation is generally non-trivial due to their off-shell nature.
These observations naturally lead to the question of whether the recursion relation in \cite{Cachazo_2022} can be improved by combining the key ideas of \cite{Cheung:2015ota} and \cite{Cachazo_2022}---namely, by using the idea of shifting Mandelstam variables and exploiting zeros of NLSM amplitudes, rather than smooth splittings, to avoid sensitivity to spacetime dimension and the introduction of off-shell currents. The recently discovered hidden zeros of NLSM amplitudes provide precisely the missing ingredient needed to realize this idea.


Over the past two years, there has been an avalanche of progress in the study of non-supersymmetric scattering amplitudes, extending geometric and combinatorial methods to theories that are closer to describing the real world \cite{Arkani-Hamed:2023lbd,Arkani-Hamed:2023mvg,Arkani-Hamed:2023swr,Arkani-Hamed:2023jry,Arkani-Hamed:2024nhp,Arkani-Hamed:2024vna,
Arkani-Hamed:2024yvu,Arkani-Hamed:2024tzl,Arkani-Hamed:2024nzc,Arkani-Hamed:2024pzc}. Among these developments, a remarkable new property of tree-level amplitudes---called the \emph{hidden zero}, referring to the vanishing of amplitudes on special loci in kinematic space---was discovered in \cite{Arkani-Hamed:2023swr}. This phenomenon was initially identified in the ${\rm Tr}(\phi^3)$, NLSM, and Yang-Mills theories, and has since been extended to a broader range of theories, including the special Galileon, Born-Infeld, and gravity \cite{Rodina:2024yfc,Bartsch:2024amu,Li:2024qfp,Zhang:2024efe,Huang:2025blb,Feng:2025ofq}. Closely related to this discovery, new factorization behaviors near hidden zero configurations---most notably the so-called 2-split---have also been intensively studied in \cite{Cao:2024gln,Cao:2024qpp,Arkani-Hamed:2024fyd,Zhang:2024iun,Zhou:2024ddy,Zhang:2024efe,Feng:2025ofq}.
In this note, we try to modify the recursion relation of \cite{Cachazo_2022} by incorporating the hidden zero as the key new ingredient.

Therefore, in this note, we propose a new BCFW-like recursion relation for tree-level NLSM amplitudes, combining the property of hidden zero with the usual property of factorization on physical poles. The hidden zero eliminates the boundary terms, while the physical-pole factorization determines the recursive part. The resulting recursion relation involves only lower-point amplitudes and does not require off-shell objects. Moreover, it is independent of the space-time dimension.


Recently, a remarkable recursive construction for ${\rm Tr}(\phi^3)$ and NLSM amplitudes, termed surface recursion, was proposed in \cite{Arkani-Hamed:2024pzc}. Surface recursion possesses several appealing features, such as greater flexibility for the choice of shifted sets and the ability to avoid spurious poles. In comparison to surface recursion, our method exhibits the following characteristics: (1) Our recursion relation treats Mandelstam variables from numerators and denominators on the same footing, whereas they are treated as distinct variables in the surface recursion; (2) Contact terms without any pole are automatically fixed by lower-point amplitudes, reflecting that vertices with different multiplicities are related by symmetry.

As demonstrations of the power of this new BCFW-like recursion relation, we apply it to recover three notable properties of tree-level NLSM amplitudes:
\begin{itemize}
\item \textbf{Adler zero:} Each tree-level NLSM amplitude vanishes when one of the external momenta is taken soft.

\item \textbf{$\delta$-shift construction of tree-level NLSM amplitudes:} Tree-level NLSM amplitudes can be obtained from ${\rm Tr}(\phi^3)$ amplitudes by performing the $\delta$-shift to ${\rm Tr}(\phi^3)$ amplitudes and taking the limit $\delta\to\infty$ after multiplying the shifted amplitudes by a monomial of $\delta$  \cite{Arkani-Hamed:2023swr,Arkani-Hamed:2024nhp}.

\item \textbf{Expansion into bi-adjoint scalar amplitudes:} Tree-level NLSM amplitudes admit an expansion into tree-level bi-adjoint scalar amplitudes, with universal coefficients \cite{Feng:2019tvb,Zhou:2023quv,Zhou:2023vzl,Zhou:2024qjh}.
\end{itemize}

The remainder of this note is organized as follows. In section~\ref{sec-BCFW-for-NLSM}, we construct the new BCFW-like recursion relation for tree-level NLSM amplitudes. Next, we consider the applications of this recursion relation. In sections~\ref{subsec-Adler}, \ref{sec-prove-delta} and~\ref{sec-expan}, we apply the recursion relation to prove the Adler zero, the $\delta$-shift construction and the universal expansion, respectively. Finally, we conclude this note with a brief discussion in section~\ref{sec-conclu}.

\section{New BCFW-like recursion relation for tree-level NLSM amplitudes}
\label{sec-BCFW-for-NLSM}

In this section, we propose a new BCFW-like recursion relation for the tree-level NLSM amplitudes, without any boundary term. This relation is constructed by combining the idea of the recursion relation in \cite{Cachazo_2022} and the recently discovered hidden zeros for NLSM amplitudes, which state that tree-level NLSM amplitudes vanish at certain loci in kinematic space.
In subsection \ref{subsec-design}, we present the explicit construction of the new recursion relation, including the general rules for shifting Mandelstam variables and modifying the NLSM amplitudes, as well as the computation of the contour integral.
Then, in subsection \ref{subsec-example}, we use the $6$-point NLSM amplitude as an example to illustrate this recursion relation.
Finally, in subsection \ref{subsec-solu-r}, we show an explicit and general type of deformations of Mandelstam variables, which can be used directly in any scenario.

The $U(N)$ NLSM under consideration is a theory of massless pions, whose Lagrangian in the Cayley parameterization \cite{Cachazo:2014xea} is
\bea
{\cal L}_{\rm NLSM}={1\over 8\lambda^2}\,{\rm Tr}(\partial_\mu {\rm U}^\dag\partial^\mu {\rm U})\,,~~\label{Lag-N}
\eea
where
\bea
{\rm U}=(\mathbb{I}+\lambda\Phi)\,(\mathbb{I}-\lambda\Phi)^{-1}\,.
\eea
Here $\mathbb{I}$ stands for the identity matrix and $\Phi=\phi_IT^I$, where $\phi_I$ is a scalar with a flavor index and $T^I$ is a generator of $U(N)$.
We are interested in the tree-level flavor-ordered partial amplitudes, denoted as $A^{\rm NLSM}_{2n}(1,\cdots,2n)$, where the coupling constants have been stripped off. Note that the non-vanishing NLSM amplitudes contain only even particles.
The Lorentz invariance forces any tree amplitude to be a function of Mandelstam variables generated by massless external momenta.
Here we clarify the notation of Mandelstam variables used in this note. The Mandelstam variable $s_{ij}$ is defined as
\bea
s_{ij}\equiv(k_i+k_j)^2=2\,k_i\cdot k_j\,,
\eea
where $k_i$ and $k_j$ are momenta of external particles $i$ and $j$. For a set of  consecutive external particles $\{i,i+1,\cdots,j-1,j\}$, the corresponding Mandelstam variable
$s_{i\cdots j}$ is defined as
\bea
s_{i\cdots j}\equiv k_{i\cdots j}^2\,,~~~{\rm with}~k_{i\cdots j}\equiv\sum_{l=i}^j\,k_l\,.
\eea
%

\subsection{Design recursion}
\label{subsec-design}

In this subsection, we explicitly construct the new recursion relation for tree-level NLSM amplitudes based on the recently discovered hidden zeros.

Instead of shifting external momenta as in the ordinary BCFW recursion relation, we consider shifting Mandelstam variables \cite{Cachazo_2022} as
\begin{align}
s_{ij}(z) = s_{ij} + z r_{ij}~~~\label{z-shift-general}
\end{align}
for $i,j\in\{1,\cdots,2n\}$, where $z\in\mathbb{C}$ is a complex variable and $r_{ij}$'s are Mandelstam-like scalar variables, which may not be the contractions of some momenta.
As will be discussed at the end of this subsection, some $r_{ij}$'s can be set to zero for specific pairs $(i,j)$.
We call such a deformation in Eq. \eref{z-shift-general} the \emph{$z$-shift} to distinguish it from the $\delta$-shift in section~\ref{sec-prove-delta}.
To keep the shifted Mandelstam variable $s_{ij}(z)$'s physical, we should require that these variables satisfy the following conditions, induced from the momentum conservation conditions and on-shell conditions,
\begin{align}
\sum_{i=1, i\neq j}^{2n} s_{ij}(z) &= 0\,,\quad{\rm for}~\forall\,j\in\{1,\cdots,2n\}\,,  \\
s_{ii}(z) &= 0, \quad \mathrm{for}~ \forall\,i\in \{1,\cdots,2n\}. \label{moment-conser}
\end{align}
Then we can conclude that the Mandelstam-like variables $r_{ij}$'s also satisfy the following momentum-conservation-like conditions and on-shell-like conditions
\begin{align}
\sum_{i=1, i\neq j}^{2n} r_{ij} &= 0\,,\quad{\rm for}~\forall\,j\in\{1,\cdots,2n\}\,,  \\
r_{ii} &= 0, \quad \mathrm{for}~ \forall\,i\in \{1,\cdots,2n\}. \label{moment-conser-r}
\end{align}
Because of the on-shell conditions for $s_{ii}$'s and on-shell-like conditions for $r_{ii}$'s, we will directly omit these variables with the same subscripts in the following for simplicity.

After the $z$-shift, the NLSM amplitude $A^{\text{NLSM}}_{2n}$ becomes a function $A^{\text{NLSM}}_{2n}(z)$ of $z$. Just as the derivation of the ordinary BCFW recursion relation, the next step is to naively apply the residue theorem to $A^{\text{NLSM}}_{2n}(z)/z$.
For now, we assume $r_{ij}\neq0$ for any $(i,j)$, then Eq. \eref{z-shift-general} implies that each $s_{ij}(z)$ is linear in $z$. Thus, $A^{\text{NLSM}}_{2n}(z)/z$ has a non-trivial residue at $z=\infty$, since the mass dimension of $A^{\text{NLSM}}_{2n}(z)$ is $2$ and then the shifted amplitude will be linear to $z$ at $z=\infty$.
Hence, in order to get rid of the boundary term, we multiply the function $A^{\text{NLSM}}_{2n}(z)/z$ by another factor $1/(1-z^2)$ as done in \cite{Cachazo_2022}, which results in the disappearance of the residue at $z=\infty$.
Then the residue theorem for $A^{\text{NLSM}}_{2n}(z)/(z(1-z^2))$ reads
\bea
A^{\rm NLSM}_{2n}={1\over 2\pi i}\,\oint_{|z|=\epsilon}\,{dz\over z}\,{A^{\rm NLSM}_{2n}(z)\over 1-z^2}\,,~~\label{contour-integral}
\eea
where $\epsilon$ is an infinitely small parameter and the residues for the poles at finite $z$ are left to analyze.

The factor $1/(1-z^2)$ in Eq. \eref{contour-integral} brings two additional poles at $z=\pm1$, which are different from those physical poles generated by the propagators. In \cite{Cachazo_2022}, the residues at $z=\pm1$ can be reduced into amputated currents by letting the poles corresponding to special splitting kinematics.
However, computing such amputated currents is still not easy, then it would be better if the poles at $z=\pm1$ don't contribute on right-hand side of Eq. \eref{contour-integral}.
Such problem can be resolved beautifully by the hidden zeros of tree NLSM amplitudes found in \cite{Arkani-Hamed:2023swr}.

\begin{figure}
    \centering
    \includegraphics[width=0.8\linewidth]{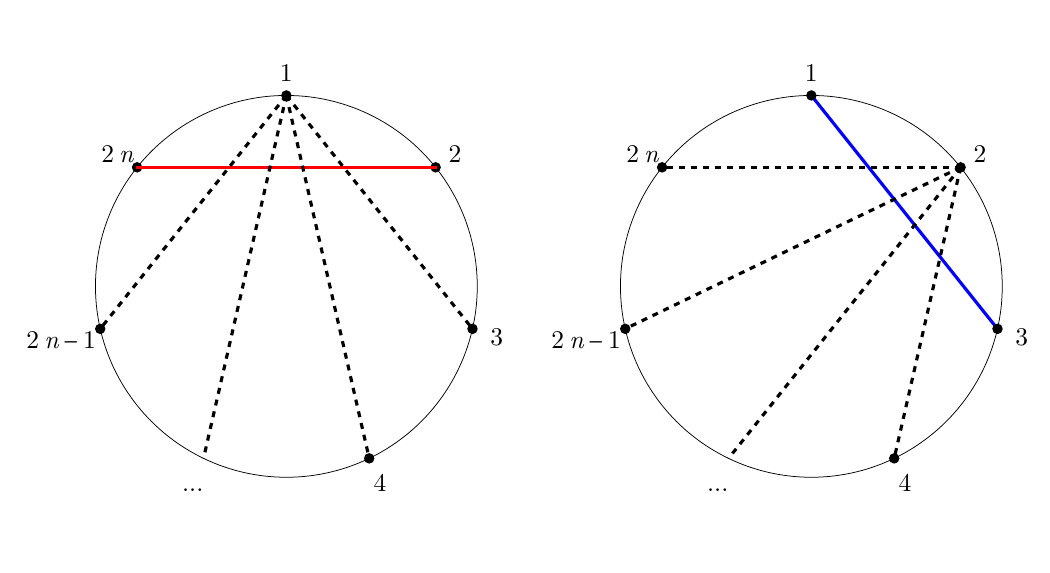}
    \caption{Illustration of the chosen hidden zeros. The left subfigure corresponds to the hidden zero $\{\tilde{i}_1, \tilde{j}_1\}=\{2n, 2\}$, the right subfigure corresponds to the hidden zero $\{\tilde{i}_2,\tilde{j}_2\}=\{1,3\}$, and the  dashed lines in each subfigure correspond to the $s_{ab}$'s in the kinematics of each hidden zero.}
    \label{fig:hidden_zeros}
\end{figure}

Hidden zeros refer to one special type of zeros of amplitudes, which arise when the Mandelstam variable $s_{ij}$'s fall into some special subspaces in the kinematic space.
More explicitly, we pick up two external legs $\tilde{i}$ and $\tilde{j}$ with $\tilde{i}<\tilde{j}$ and $\tilde{j}\neq \tilde{i}+1$, cyclically split the external legs into two sets $A=\{\tilde{i}+1,\cdots,\tilde{j}-1\}$ and $B=\{\tilde{j}+1,\cdots,\tilde{i}-1\}$, then $A^{\rm NLSM}_{2n} \rightarrow 0$ when
\bea
s_{ab}=0\,,~\forall~a\in A\,,~b\in B.~~~~\label{kinematic-condi-zero}
\eea
Through this note, we will refer to Eq. \eref{kinematic-condi-zero} as the kinematics for the hidden zero $\{\tilde{i},\tilde{j}\}$, as illustrated in Fig. \ref{fig:hidden_zeros}.
Note that the shifted amplitude $A^{\rm NLSM}_{2n}(z)$ also vanishes, when the shifted variable $s_{ij}(z)$'s take the analogous kinematics of hidden zero $\{\tilde{i},\tilde{j}\}$, i.e., $s_{ab}(z)=0$ for all $a\in A, b\in B$.
In order to let the kinematics of the shifted Mandelstam variable $s_{ij}(z)$'s at $z=\pm1$ be the analogous kinematics of two hidden zeros of $\{\tilde{i}_1,\tilde{j}_1\}$ and $\{\tilde{i}_2,\tilde{j}_2\}$ respectively, which correspond to $\{A_1,B_1\}$ and $\{A_2,B_2\}$, we should require
\begin{align}
r_{a_1b_1}&=-s_{a_1b_1}\,,\,~~~~{\rm for}~a_1\in A_1\,,~b_1\in B_1, \notag\\
r_{a_2b_2}&=s_{a_2b_2}\,,\,~~~~~~{\rm for}~a_2\in A_2\,,~b_2\in B_2.~~\label{realize-zero}
\end{align}
If we make some choices of $r_{ij}$'s as in Eq. \eref{realize-zero}, the contour integral in Eq. \eref{contour-integral} around poles $z=\pm1$ forces the kinematics of the shifted Mandelstam variable $s_{ij}(z)$'s at $z=\pm1$ reduce to the kinematics of two hidden zeros of $\{\tilde{i}_1,\tilde{j}_1\}$ and $\{\tilde{i}_2,\tilde{j}_2\}$, then the residues at $z=\pm1$ vanish.

Note that the kinematics corresponding to two hidden zeros, ${\tilde{i}_1,\tilde{j}_1}$ and ${\tilde{i}_2,\tilde{j}_2}$, should not overlap. In other words, any equation $s_{ab}=0$ for specific $a$ and $b$ cannot appear simultaneously in the kinematic conditions of both hidden zeros at $z=\pm 1$. Otherwise, the choices of $r_{a_1b_1}$ and $r_{a_2b_2}$ in Eq.~\eref{realize-zero} would become inconsistent unless the $z$-shift in Eq.~\eref{z-shift-general} is modified. Therefore, to maintain consistency, such overlaps must be avoided.
As a simple example, consider the $6$-point amplitude $A^{\rm NLSM}_6(1,2,3,4,5,6)$. Suppose $\{\tilde{i}_1,\tilde{j}_1\}=\{1,4\}$ and $\{\tilde{i}_2,\tilde{j}_2\}=\{2,5\}$. The first zero at $z=1$ requires $s_{36}(1)=0$, while the second zero requires $s_{36}(-1)=0$. Consequently, the shifted invariant has to be $s_{36}(z)=(1-z^2)s_{36}$, which breaks the desired linear dependence on $z$.

Fortunately, two hidden zeros without overlapping always exist, for example, $\{\tilde{i}_1,\tilde{j}_1\}=\{2n,2\}$ and $\{\tilde{i}_2,\tilde{j}_2\}=\{1,3\}$ as illustrated in Fig. \ref{fig:hidden_zeros}.
After choosing two hidden zeros $\{\tilde{i}_1,\tilde{j}_1\}$ and $\{\tilde{i}_2,\tilde{j}_2\}$ and setting Eq. \eref{realize-zero},  the contour integral in Eq. \eref{contour-integral} can be computed as
\begin{align}
\label{eq:BCFW_recursion_hidden_zero}
A^{\text{NLSM}}_{2n} =& -\sum_{z^*\neq0,\pm1,\infty}\,{\rm Res}_{z=z^*}\,{A^{\rm NLSM}_{2n}(z)\over z\,(1-z^2)} \notag\\
=& \sum_{z^*: S(z^*)=0} A_{2n_1}^{\text{NLSM}}(z^*) \frac{1}{[1-(z^*)^2]S} A_{2n+2-2n_1}^{\text{NLSM}}(z^*)\,,
\end{align}
where $n_1<n$, $S$'s are planar Mandelstam invariants and the summation is taken over all finite $z^*$ satisfying $s(z^*)=0$. This on-shell recursion relation is the first main result of this paper.
Compared with the recursion relation in \cite{Cachazo_2022},
the new recursion relation in Eq. \eref{eq:BCFW_recursion_hidden_zero} resembles the ordinary BCFW recursion relations, reducing the computation of a higher-point amplitude into the computations of on-shell lower-point amplitudes.

\begin{figure}
    \centering
    \includegraphics[width=0.45\linewidth]{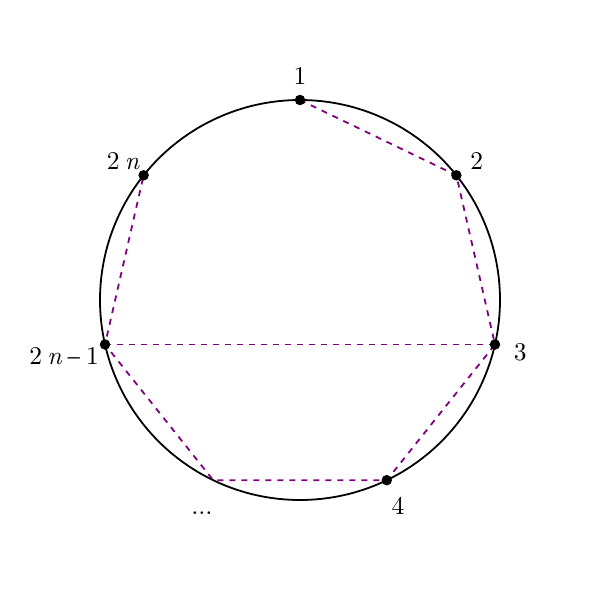}
    \caption{Illustration of the set of $r_{ij}$'s solved by the momentum-conservation-like conditions in Eq. \eref{moment-conser-r}, and the purple dashed lines corresponds the $r_{ij}$'s. }
    \label{fig:momentum_conservation_r_ij}
\end{figure}

Through the $z$-shift defined in Eq. \eref{z-shift-general}, we introduce a total of $n(2n-1)$ auxiliary Mandelstam-like variables $r_{ij}$. While the momentum conservation conditions fix $2n$ of these variables, and the kinematics of the two hidden zeros in Eq. \eref{realize-zero} determine additional $2(2n-3)$ variables, there remain $(n-2)(2n-3)$ undetermined $r_{ij}$'s. Because the final recursion relation in Eq. \eref{eq:BCFW_recursion_hidden_zero} must be independent of these auxiliary variables, we can directly let $(n-2)(2n-3)$ $r_{ij}$'s equal to $0$.
However, the previous counting for the power of $z$ in the large $z$ limit---which motivates the factor $1/(1-z^2)$ in Eq. \eref{contour-integral}---is based on assuming each Mandelstam variable to be linear in $z$, and the above $r_{ij}=0$ breaks such linearity. The solution is to choose $r_{ij}$ such that every denominator of physical propagator remains linear in $z$. This choice thereby prevents a non-vanishing boundary term in Eq. \eref{contour-integral} when $r_{ij}=0$, since the power of $z$ in the limit $z\to\infty$ cannot be increased.
In subsection \ref{subsec-solu-r}, we will provide an explicit choice of $(n-2)(2n-3)$ $r_{ij}$'s for the general $2n$-point NLSM amplitudes, as illustrated in Fig. \ref{fig:momentum_conservation_r_ij}.

\subsection{Example: $6$-point NLSM amplitude}
\label{subsec-example}

In this subsection, we take the $6$-point NLSM amplitude as an example to illustrate the computation procedure of $A^{\rm NLSM}_6(1,\cdots,6)$ via the recursive relation in Eq. \eref{eq:BCFW_recursion_hidden_zero}.

Let two hidden zeros of $\{\tilde{i}_1,\tilde{j}_1\}=\{6,2\}$ and $\{\tilde{i}_2,\tilde{j}_2\}=\{1,3\}$ correspond to the poles at $z=\pm 1$ respectively.
According to Eq. \eref{realize-zero}, $6$ auxiliary Mandelstam-like variables should be chosen as
\bea
& &r_{13}=-s_{13}\,,~~r_{14}=-s_{14}\,,~~r_{15}=-s_{15}\,,\nn
& &r_{24}=s_{24}\,,~~~~r_{25}=s_{25}\,,~~~~r_{26}=s_{26}\,.~~~~\label{solur-6pt1}
\eea
At the same time, we should solve the momentum conservation conditions and obtain the following solution,
\bea
r_{12}=s_{13}+s_{14}+s_{15}\,,&~~&r_{23}=-\big(s_{13}+s_{14}+s_{15}\big)-\big(s_{24}+s_{25}+s_{26}\big)\,,\nn
r_{34}=s_{13}+s_{14}+s_{25}\,,&~~&r_{45}=-s_{13}-s_{24}-s_{25}\,,\nn
r_{56}=-s_{26}\,,~~~~~~~~~~~~&~~&r_{35}=s_{13}+s_{15}+s_{24}+s_{26}\,,~~~~\label{solur-6pt2}
\eea
where we have chosen $3$ $r_{ij}$'s and let them be $0$ as
\bea
r_{16}=r_{36}=r_{46}=0\,.
~~~~\label{solur-6pt3}
\eea
Because of the facts that the NLSM vertices are connected with even lines and the NLSM amplitudes are flavor-ordered, then $A^{\rm NLSM}_6$ has propagators $s_{123}$, $s_{234}$ and $s_{345}$.
We can verify that under the $z$-shift with $r_{ij}$ being given in Eq. \eref{solur-6pt1}, \eref{solur-6pt2} and \eref{solur-6pt3}, all three planar Mandelstam invariants $\{ s_{123}, s_{234}, s_{345}\}$ are linear in $z$.

To apply the BCFW-like recursion relation in Eq.~\eref{eq:BCFW_recursion_hidden_zero} to the $6$-point NLSM amplitude, the explicit expression of the $4$-point NLSM amplitude is required as the starting point.
The $4$-point NLSM amplitude can be determined by the following procedure.
Since the $4$-point NLSM amplitude doesn't contain any propagator and has mass dimension $2$, then the amplitude $A^{\rm NLSM}_4(1,2,3,4)$ must be a linear combination of Mandelstam invariants $s$, $t$ and $u$.
Furthermore, the flavor-ordered amplitude $A^{\rm NLSM}_4(1,2,3,4)$ satisfies the cyclic symmetry, then $A^{\rm NLSM}_4(1,2,3,4)\propto s+t$ or $A^{\rm NLSM}_4(1,2,3,4)\propto u$.
Because of $s+t+u=0$, therefore the amplitude can be uniquely determined as $A^{\rm NLSM}_4(1,2,3,4)=s+t$ up to an irrelevant constant coefficient.

Next, we show the explicit computation procedure of the BCFW-like recursion relation in Eq.~\eref{eq:BCFW_recursion_hidden_zero}. The first contribution comes from $s_{123}(z^*)=0$, which corresponds to the pole $z^*=s_{123}/(s_{13}-s_{12}-s_{23})$, then the residue is given as
\begin{align}
{\rm Res}_{z^*: s_{123}(z^*)=0} \frac{A_{6}^{\mathrm{NLSM}}}{z(1-z^2)}
&=[s_{12}(z^*)+s_{23}(z^*)] \frac{1}{s_{123}[1-(z^*)^2]} [s_{45}(z^*)+s_{56}(z^*)] \notag\\
&=\frac{(s_{12}+s_{23})(s_{45}+s_{56})}{s_{123}} + \frac{1}{2}(s_{13}-s_{12}-s_{23}-s_{45}-s_{56}).~~\label{Residue1}
\end{align}
Then we consider the second contribution from $s_{234}(z^*)=0$, which is given as
\begin{align}
{\rm Res}_{z^*:s_{234}(z^*)=0}\frac{A_{6}^{\mathrm{NLSM}}}{z(1-z^2)}
&=[s_{23}(z^*)+s_{34}(z^*)] \frac{1}{s_{234}[1-(z^*)^2]} [s_{56}(z^*)+s_{61}(z^*)] \notag\\
&=\frac{(s_{23}+s_{34})(s_{61}+s_{56})}{s_{234}} + s_{24}-s_{61}-s_{56}.~~\label{Residue2}
\end{align}
The final contribution comes from $s_{345}(z^*)=0$,
\begin{align}
{\rm Res}_{z^*:s_{345}(z^*)=0}\frac{A_{6}^{\mathrm{NLSM}}}{z(1-z^2)} =\,\frac{(s_{12}+s_{16})(s_{34}+s_{45})}{s_{345}} + \frac{1}{2}(s_{14}-s_{24}-s_{34}-s_{45}).~~\label{Residue3}
\end{align}
Combining the results in Eq. \eref{Residue1}, \eref{Residue2} and \eref{Residue3} and making some simplification, we get the following $6$-point NLSM amplitude
\begin{align}
A^{\rm NLSM}_6=&\frac{(s_{12}+s_{23})(s_{45}+s_{56})}{s_{123}} + \frac{(s_{23}+s_{34})(s_{61}+s_{56})}{s_{234}}+\frac{(s_{12}+s_{16})(s_{34}+s_{45})}{s_{345}} \notag\\
& - \big(s_{12}+s_{23}+s_{34}+s_{45}+s_{56}+s_{61}\big)\,,
\end{align}
which coincides with the known result.

\subsection{$r_{ij}$ for general $2n$-point amplitude}
\label{subsec-solu-r}

As stated in the last paragraph of Subsection~\ref{subsec-design}, after fixing $2(2n - 3)$ of the $r_{ij}$ parameters using the kinematics of the two hidden zeros, we are left with $n(2n - 1)$ $r_{ij}$ variables subject to $2n$ momentum conservation conditions. Therefore, there remains considerable freedom in choosing a subset of the $r_{ij}$'s, which can be set to zero.
In this subsection, we present a general choice for the $r_{ij}$'s, which generalizes the specific cases shown in Eq. \eref{solur-6pt1}, \eref{solur-6pt2}, and \eref{solur-6pt3}, and is illustrated in Fig. \ref{fig:hidden_zeros}.

We let the kinematics of two hidden zeros $\{\tilde{i}_1,\tilde{j}_1\}=\{2n,2\}$ and $\{\tilde{i}_2,\tilde{j}_2\}=\{1,3\}$ correspond to $z=\pm 1$ respectively, which determines $2(2n-3)$ $r_{ij}$'s as
\begin{align}
r_{1i} &= -s_{1i}\,, \quad \ \text{for}~ i\in \{3,4,\cdots,2n-1\}, \notag \\
r_{2j} &= s_{2j}\,,~~~ \quad \text{for}~ j\in \{4,5,\cdots,2n\}.~~\label{solur-1}
\end{align}
The momentum conservation equations in Eq. \eref{moment-conser-r} allow us to determine $2n$ $r_{ij}$'s, for example $r_{3,2n-1}$ and $r_{i,i+1}$ with $i=1,\cdots,2n-1$, as visualized by the dashed lines in Fig. \ref{fig:momentum_conservation_r_ij}.\footnote{One may wonder if the variable $r_{3,2n-1}$ can be replaced by $r_{2n,1}$ which makes the solution or the graph represented by the dashed lines in Fig. \ref{fig:momentum_conservation_r_ij} to be more symmetric. However, for the latter choice, there does not exist a consistent solution of $r_{ij}$'s.}
The solution to the momentum conservation equations can be found as
\begin{align}
& r_{12} = \sum_{a=3}^{2n-1}s_{1a}\,, \quad
r_{23} = -\sum_{a=3}^{2n-1}s_{1a} - \sum_{b=4}^{2n}s_{2b}\,, \quad
r_{34} = s_{13} + \sum_{a=2}^{n-1}s_{1,2a} + \sum_{b=3}^n s_{2,2b-1}\,, \notag\\
& r_{3,2n-1} = \sum_{a=2}^n s_{1,2a-1} + \sum_{b=2}^n s_{2,2b}\,, \quad
r_{2n-1,2n} = -s_{2,2n}\,.~~~~\label{solur-2}
\end{align}
and
\begin{align}
r_{2k,2k+1} &= -\sum_{a=2}^k s_{1,2a-1} - \sum_{b=k+1}^{n-1}s_{1,2b} - \sum_{c=2}^k s_{2,2c} - \sum_{d=k+1}^n s_{2,2d-1}, \quad \text{for}~k=2,\cdots,n-1\,, \notag\\
r_{2k-1,2k} &= \sum_{a=2}^k s_{1,2a-1} + \sum_{b=k}^{n-1}s_{1,2b} + \sum_{c=2}^{k-1} s_{2,2c} + \sum_{d=k+1}^{n} s_{2,2d-1}, \quad \text{for}~k=3,\cdots,n-1\,.~~~~\label{solur-3}
\end{align}
Finally, all remaining $r_{ij}$'s are set to be $0$.
It is straightforward to check that under the $z$-shift in Eq. \eref{z-shift-general} with $r_{ij}$'s being given in Eq. \eref{solur-1}, \eref{solur-2} and \eref{solur-3},
all planar Mandelstam invariants $s_{i\ldots j}(z)$ with $i$ and $j$ being simultaneously odd or even are linear in $z$, thus all propagators in the shifted amplitude $A_{2n}^{\mathrm{NLSM}}(z)$ are linear in $z$ and the residue of $A_{2n}^{\mathrm{NLSM}}(z)/(1-z^2)$ at $z=\infty$ vanishes.

\section{Applications}
In subsection \ref{subsec-design}, we pointed out that the new BCFW-like recursion relation given in Eq. \eref{eq:BCFW_recursion_hidden_zero} possesses several favorable features: it avoids complications from boundary terms and relies solely on lower-point amplitudes as input.
These advantageous properties make the recursion relation a powerful tool for establishing many important results about NLSM amplitudes.

\subsection{Application \RNum{1}: Adler zero}
\label{subsec-Adler}
In this subsection, we show the first of three applications: proving the Adler zero for the NLSM amplitude, which states that each tree NLSM amplitude vanishes when one external momentum is soft.

Now we consider using the recursion relation in Eq. \eref{eq:BCFW_recursion_hidden_zero} to prove the Adler zero with the external momentum $k_1$ being soft: more explicitly, $A_{2n}^{\mathrm{NLSM}}=0$ when $k_1 \to 0$ or $s_{1i}\to 0$ for $i=2,\cdots,2n$.  According to the analysis in the previous section, we know that $A_{4}^{\mathrm{NLSM}}\propto s+t$, then $A_{4}^{\mathrm{NLSM}}$ vanishes when $s_{1i}=0$ for $i=2,3,4$.
Assuming the Adler zero of external momentum $k_1$ exists for $A_{2n-2}^{\mathrm{NLSM}}$, then we try to prove Adler zero for $A_{2n}^{\mathrm{NLSM}}$ using the BCFW-like recursion relation with two hidden zeros $\{\tilde{i}_1,\tilde{j}_1\}=\{2n,2\}$ and $\{\tilde{i}_2,\tilde{j}_2\}=\{1,3\}$.
Using the solution of $r_{ij}$'s provided in the previous subsection, we see that each $s_{1i}(z)$ also behaves as
$s_{1i}(z)\to 0$, since $r_{1i}\to 0$.
Due to recursive assumption, every sub-amplitudes in Eq. \eref{eq:BCFW_recursion_hidden_zero} including the external leg labelled $1$ vanishes, thus the full $2n$-point amplitude on the left hand side of Eq. \eref{eq:BCFW_recursion_hidden_zero} also vanishes.
So we have proved the Adler zero of $k_1$ for $2n$-point NLSM amplitude.
Adler zeros for other external momenta can be obtained via the cyclic symmetry.

It is worthwhile to compare the Adler zero with the hidden zero.
For Adler zero, without loss of generality, we assume $k_1$ is soft, i.e., $k_1\to 0$, then the associated Mandelstam variables $s_{1i}$ behave as
\bea
s_{1i}\,\to\,0\,,~~~~{\rm for}~\forall\,i\in\{2,\cdots,2n\}\,.~~~~\label{1-soft-kine}
\eea
Thus Adler zero says that for the kinematics $\{s_{1i}=0, i=2,\cdots,2n\}$, the NLSM amplitude vanishes. As for hidden zero, for example $\{\tilde{i},\tilde{j}\}=\{2n,2\}$ as in Fig.\ref{fig:hidden_zeros}, then the kinematics of hidden zero kinematics in \eref{kinematic-condi-zero} is
\bea
s_{1j}=0\,,~~~~{\rm for}~\forall\,j\in\{3,\cdots,2n-1\}.~~\label{1-hidden0}
\eea
Comparing Eq. \eref{1-soft-kine} and Eq. \eref{1-hidden0}, the kinematics of Adler zero in Eq. \eref{1-soft-kine} contain two extra conditions, i.e., $s_{12}=0$ and $s_{1,2n}=0$. Hence, the hidden zero is more general than the Adler zero since it imposes a weaker kinematic condition on NLSM amplitudes. Therefore, factorization on physical poles together with the hidden zero serves as an enhanced set of principles that fully determines NLSM amplitudes.

\subsection{Application \RNum{2}: the $\delta$-shift construction of NLSM amplitude}
\label{sec-prove-delta}
In this subsection, we use the BCFW-like recursion relation Eq. \eref{eq:BCFW_recursion_hidden_zero} to prove
a remarkable discovery that the NLSM amplitudes can be constructed from ${\rm Tr}(\phi^3)$ amplitudes via the $\delta$-shift.
In subsection \ref{subsec-delta}, we first briefly review the $\delta$-shift construction. Then, we present a recursive proof of this $\delta$-shift construction in subsection \ref{subsec-prove-delta}.

\subsubsection{Constructing NLSM amplitudes from ${\rm Tr}(\phi^3)$ amplitudes via $\delta$-shift}
\label{subsec-delta}

In \cite{Arkani-Hamed:2023swr,Arkani-Hamed:2024nhp}, the authors found that the NLSM amplitudes can be constructed from ${\rm Tr}(\phi^3)$ amplitudes by applying the $\delta$-shift and extracting the leading contribution in the large $\delta$ limit. More explicitly, the $\delta$-shift for Mandelstam variables is given as
\begin{align}
\left\{ \begin{array}{cc}
s_{i\cdots j} \to s_{i\cdots j} - \delta\,,~~~~ & \text{if $i$ is odd and $j$ is even,} \\
s_{i\cdots j} \to s_{i\cdots j} + \delta\,, ~~~~& \text{if $i$ is even and $j$ is odd.}
\end{array} \right.~~\label{delta-shift-original}
\end{align}
Note that the mass dimension of $\delta$ is the same as Mandelstam variables, i.e., $2$.
After performing this shift, the NLSM amplitude can be obtained\footnote{The limit presented here differs from that in \cite{Arkani-Hamed:2023swr,Arkani-Hamed:2024nhp} by a minus sign, due to our choice of the expression of $4$-point NLSM amplitude.}
\begin{align}
\label{eq:nlsm_phi_cube_shift}
A_{2n}^{\te{NLSM}} = -\lim_{\delta \rightarrow \infty} \delta^{2n-2}\,A_{2n}^{{\rm Tr}(\phi^3)}(\delta)\,.
\end{align}

First, let us consider the $4$-point amplitude ${\cal A}^{\rm NLSM}_4$ as the simplest example to illustrate the above construction Eq. \eref{eq:nlsm_phi_cube_shift}, which also serve as a starting point of our recursive proof of Eq. \eref{eq:nlsm_phi_cube_shift}.
The shifted $4$-point $\Tr(\phi^3)$ amplitude is
\begin{align}
A_{4}^{{\rm Tr}(\phi^3)}(\delta) = \frac{1}{s_{12}-\delta} + \frac{1}{s_{23}+\delta}
= \frac{1}{\delta}\left[ -\sum_{n=0}^{\infty} \left(\frac{s_{12}}{\delta}\right)^n + \sum_{n=0}^{\infty}(-1)^n \left(\frac{s_{23}}{\delta}\right)^n \right]
=-\frac{1}{\delta^2}(s_{12}+s_{23}) + {\cal O}(\delta^{3})^{-1}\,,~~\label{4p-delta-expan}
\end{align}
then the coefficient of the leading term gives rise to $4$-point NLSM amplitude correctly,
\begin{align}
\label{eq:4_nlsm_phi_cube}
A_{4}^{\te{NLSM}} = -\lim_{\delta \rightarrow \infty} \delta^{2} A_{4}^{{\rm Tr}(\phi^3)}(\delta) = s_{12}+s_{23}.
\end{align}

For latter convenience, we introduce the following alternative $\delta$-shift which acts on $2$-point Mandelstam variables,
\begin{align}
\left\{ \begin{array}{cc}
s_{i,i+1} \to s_{i,i+1} - \delta,& \quad\text{if $i$ is odd,} \\
s_{i,i+1} \to s_{i,i+1} + \delta,& \quad\text{if $i$ is even.}
\end{array} \right.~~\label{delta-shift-our}
\end{align}
When the total number of external particles is even, it is straightforward to verify that Eq. \eref{delta-shift-our} is exactly equivalent to Eq. \eref{delta-shift-original}.
We also replace the large $\delta$ limit in Eq. \eref{eq:nlsm_phi_cube_shift} by a contour integral as
\bea
A_{2n}^{\rm NLSM} ={1\over 2\pi i}\,\oint_{\delta=\infty}\,{d\delta\over\delta}\, \delta^{2n-2}\,A_{2n}^{{\rm Tr}(\phi^3)}(\delta)\,.~~\label{to-prove-delta}
\eea
It is easy to see that the operation of taking contour integration ${1\over 2\pi i}\oint_{\delta=\infty}{d\delta}\delta^{2n-3}$ can extract the coefficient of $1/\delta^{2n-2}$, thus applying the contour integral to $A_{4}^{{\rm Tr}(\phi^3)}(\delta)$, we get the correct $4$-point NLSM amplitude just as Eq. \eref{eq:4_nlsm_phi_cube}.
However, the contour integral in Eq. \eref{to-prove-delta} has an advantage over taking the the large $\delta$ limit of Eq. \eref{eq:nlsm_phi_cube_shift}: although we know the leading term of $A_{4}^{{\rm Tr}(\phi^3)}(\delta)$ in the large $\delta$ limit has the form $1/\delta^{2}$, in general it is difficult to show whether the leading term of $A_{2n-2}^{{\rm Tr}(\phi^3)}(\delta)$ has the order $-(2n-2)$, thus taking the the large $\delta$ limit of Eq. \eref{eq:nlsm_phi_cube_shift} may results in a divergence, while the contour integral in Eq. \eref{to-prove-delta} only extracts the coefficient of $1/\delta^{2n-2}$ and then will not encounter divergence.
This feature of the form of contour integral will become useful when we try to prove Eq. \eref{to-prove-delta} recursively in the next subsection.

\subsubsection{Recursive proof}
\label{subsec-prove-delta}

In this subsection, we use the BCFW-like recursion in Eq. \eref{eq:BCFW_recursion_hidden_zero} to prove the $\delta$-shift construction of NLSM amplitudes in Eq. \eref{to-prove-delta}. Explicitly, our purpose is to show
\bea
{1\over 2\pi i}\,\oint_{\delta=\infty}\,{d\delta\over\delta}\, \delta^{2n-2}\,A_{2n}^{{\rm Tr}(\phi^3)}(\delta)=\sum_{z^*: S(z^*)=0} A_{2n_1}^{\text{NLSM}}(z^*) \frac{1}{[1-(z^*)^2]S} A_{2n+2-2n_1}^{\text{NLSM}}(z^*)\,.~~\label{goal-delta}
\eea
Note that $A_{2n}^{\rm NLSM}$ in Eq. \eref{to-prove-delta} has been replaced by the recursion relation in Eq. \eref{eq:BCFW_recursion_hidden_zero}.
Since the $\delta$-shift construction in the $4$-point case has been proved in the previous subsection, then according to the spirit of recursive proof, we assume that such $\delta$-shift construction is valid for NLSM amplitudes with the number of external lines no larger than $2n-2$.

In the next, we prove Eq. \eref{goal-delta} for $2n$-point NLSM amplitude. We apply the operations for obtaining the BCFW-like recursion relation to the left hand side (l.h.s) of Eq. \eref{goal-delta},
\begin{align}
{1\over 2\pi i}\,\oint_{\delta=\infty}\,{d\delta\over\delta}\, \delta^{2n-2}\,A_{2n}^{{\rm Tr}(\phi^3)}(\delta) =& -{1\over 4\pi ^2}\,\oint_{|z|=\epsilon}\, {dz\over z}\,{ 1\over 1-z^2}\,\Big[\oint_{\delta=\infty}\,{d\delta\over\delta}\, \delta^{2n-2}\,A_{2n}^{{\rm Tr}(\phi^3)}(\delta)\Big]_{\text{$z$-shift}}\nn
=& -\sum_{z^*\ne0}\,{\rm Res}_{z=z^*}\,{ 1\over z(1-z^2)}\,\Big[{1\over 2\pi i}\,\oint_{\delta=\infty}\,{d\delta\over\delta}\, \delta^{2n-2}\,A_{2n}^{{\rm Tr}(\phi^3)}(\delta)\Big]_{\text{$z$-shift}}\nn
=& -\sum_{z^*\ne0}\,{\rm Res}_{z=z^*}\,{ 1\over z(1-z^2)}\,\Big[{1\over 2\pi i}\,\oint_{\delta=\infty}\,{d\delta\over\delta}\, \delta^{2n-2}\,A_{2n}^{{\rm Tr}(\phi^3)}(\delta,z)\Big]\,,~~\label{reform-delta}
\end{align}
where $z$-shift is chosen as in Eq. \eref{realize-zero}, letting the kinematics of the shifted Mandelstam variable $s_{ij}(z)$'s at $z=\pm1$ to satisfy the kinematics of two hidden zeros of $\{\tilde{i}_1,\tilde{j}_1\}=\{2n, 2\}$ and $\{\tilde{i}_2,\tilde{j}_2\}=\{1, 3\}$.
Here $A_{2n}^{{\rm Tr}(\phi^3)}(\delta,z)$ in the last step is defined as
\bea
A_{2n}^{{\rm Tr}(\phi^3)}(\delta,z):=\Big[A_{2n}^{{\rm Tr}(\phi^3)}(\delta)\Big]_{z-{\rm shift}}\,,~~\label{delta-z}
\eea
where $z$-shift is applied after $\delta$-shift. Since $z$-shift does not act on $\delta$, then $A_{2n}^{{\rm Tr}(\phi^3)}(\delta,z)$ doesn't contain such factor $z\delta$.
Applying the $z$-shift to this residue is equal to the result of taking the residue of the $\delta^{2n-3}A_{2n}^{{\rm Tr}(\phi^3)}(\delta,z)$, thus the last line of Eq. \eref{reform-delta} holds. Notice that if we use the large $\delta$ limit formula instead of the contour integral formula, the analogue of last step of Eq. \eref{reform-delta} can not be ensured, due to the potential divergence.

Now we need to analyze the residues for different types of poles in the last line of Eq. \eref{reform-delta}.

\textbf{Residue for pole $z=\infty$~}
The residue at $z=\infty$ vanishes and the reason is simple as following. Since the function of Mandelstam variables arising from the contour integral around $\delta=\infty$ has mass dimension $2$, thus the $z$-shift indicates that the leading term of the Laurent series of this function at $z=\infty$ must be proportional to $z$. Then the pre-factor $1/z(1-z^2)$ let the leading term be proportional to $z^{-2}$, so the residue at $z=\infty$ vanishes.

\textbf{Residues for pole $z=\pm1$~}
Since there doesn't exist such factor $z\delta$ in $A_{2n}^{{\rm Tr}(\phi^3)}(\delta,z)$, then we can equivalently reinterpret $A_{2n}^{{\rm Tr}(\phi^3)}(\delta,z)$ as the result of performing the $\delta$-shift to the $z$-shifted amplitude $A_{2n}^{{\rm Tr}(\phi^3)}(z)$, i.e.,
\begin{align}
A_{2n}^{{\rm Tr}(\phi^3)}(\delta,z)=A_{2n}^{{\rm Tr}(\phi^3)}(z, \delta)=\Big[A_{2n}^{{\rm Tr}(\phi^3)}(z)\Big]_{\delta-{\rm shift}},~~\label{alter-understand}
\end{align}
where $\delta$-shift is applied to the $z$-shifted Mandelstam variables $s_{ij}(z)$ following the similar rules as Eq. \eref{delta-shift-our}, i.e.,
\begin{align}
\left\{ \begin{array}{cc}
s_{i,i+1}(z) \to s_{i,i+1}(z) - \delta\,, ~~~~& \text{if $i$ is odd,} \\
s_{i,i+1}(z) \to s_{i,i+1}(z) + \delta\,,~~~~ & \text{if $i$ is even.}
\end{array} \right.~~\label{delta-shift-1}
\end{align}
Next, we can exploit the fact that the $\mathrm{Tr}(\phi^3)$ amplitude exhibits hidden zeros at the same kinematics of Eq. \eref{kinematic-condi-zero} in kinematic space of the NLSM amplitude \cite{Arkani-Hamed:2023swr} to conclude that the residues at $z=\pm1$ vanish.
Let us take the pole at $z=1$ as an example. Since the hidden zero at $z=1$ forces $A_{2n}^{{\rm Tr}(\phi^3)}(z)$ to be $0$, then $A_{2n}^{{\rm Tr}(\phi^3)}(z)$ can be expanded as
\bea
A_{2n}^{{\rm Tr}(\phi^3)}(z)=\sum_{a_1\in A_1,b_1\in B_1}\,s_{a_1b_1}(z)\,B_{a_1b_1}(z)\,,~~\label{expan-s}
\eea
where each shifted Mandelstam variable $s_{a_1b_1}(z)$ vanishes at $z=1$. In Appendix \ref{sec-Appen}, we will give an argument for the existence of the above expansion Eq. \eref{expan-s}.
According to Eq. \eref{kinematic-condi-zero}, each pair of legs $a_1$ and $b_1$ are separated by $\tilde{i}_1$ and $\tilde{j}_1$ and then cannot be consecutive, i.e., $b_1\neq a_1\pm1$, then $\delta$-shift following the rules of Eq. \eref{delta-shift-our} does not affect these $s_{a_1b_1}(z)$, then
\bea
A_{2n}^{{\rm Tr}(\phi^3)}(z, \delta)=\sum_{a_1\in A_1,b_1\in B_1}\,s_{a_1b_1}(z)\,\Big[B_{a_1b_1}(z)\Big]_{\delta-{\rm shift}}\,.
\eea
Thus $A_{2n}^{{\rm Tr}(\phi^3)}(z, \delta)$ vanishes when $s_{a_1b_1}(z)=0$ at $z=1$, i.e., $A_{2n}^{{\rm Tr}(\phi^3)}(z, \delta)$ has the hidden zero corresponding to $\{\tilde{i}_1,\tilde{j}_1\}=\{2n, 2\}$.  Hence, the residue in Eq. \eref{reform-delta} vanishes at $z=1$.
The case is the same for $z=-1$.

\begin{figure}
    \centering
    \includegraphics[width=0.52\linewidth]{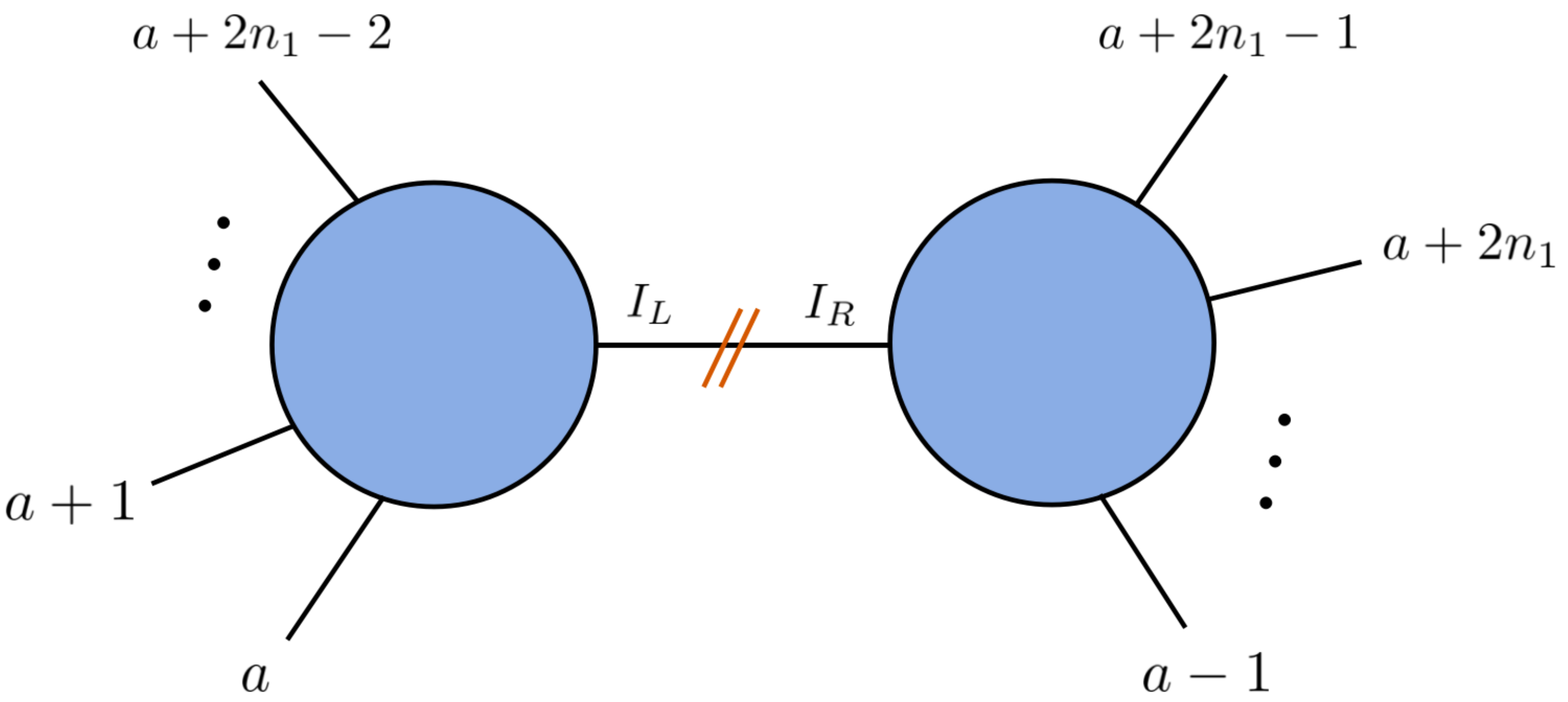}
    \caption{Illustration of the case of $|L|$ being odd. Here the internal line $S_L(z)$ is cut and generated two external lines $I_L$ and $I_R$. }
    \label{fig:recursion_cut}
\end{figure}

\textbf{Residues for other poles~}
Now we analyze the locations of remaining poles from
\bea
{1\over 2\pi i}\,\oint_{\delta=\infty}\,{d\delta\over\delta}\, \delta^{2n-2}\,A_{2n}^{{\rm Tr}(\phi^3)}(\delta,z)\,,~~\label{contour-inte-delta}
\eea
by considering the deformed amplitude $A_{2n}^{{\rm Tr}(\phi^3)}(\delta,z)$. The amplitude $A_{2n}^{{\rm Tr}(\phi^3)}(\delta,z)$ consists solely of scalar propagators $1/S_L(z,\delta)$ where
\bea
S_L(z,\delta)=S_L(z)\Big|_{\delta-{\rm shift}}\,,~~~~~~~~S_L(z)=\sum_{a,b\in L}\,s_{ab}(z)\,.
\eea
In the above, $L$ denotes the set of external legs on the left of this propagator, as illustrated in Fig. \ref{fig:recursion_cut}. These propagators can be naturally classified into two sectors, with $|L|$ being even or odd respectively, where $|L|$ stands for the cardinality of $L$. When $|L|$ is even, then
the rules for $\delta$-shift in \eref{delta-shift-1} indicates
\bea
S_L(z,\delta)=S_L(z)\,\pm\,\delta\,,
\eea
thus one can expand the corresponding propagator around $\delta\to\infty$ as in \eref{4p-delta-expan}, namely
\bea
{1\over S_L(z,\delta)}=\pm\,{1\over\delta}\,\Big[\sum_{n=0}^\infty\,\Big({\mp\,S_L(z)\over\delta}\Big)^n\Big]\,.
\eea
In the above expansion, the dependence on $z$ only appears in numerators which are independent of $\delta$, thus these propagators will never contribute any pole after performing the contour integration \eref{contour-inte-delta}.

If $|L|$ is odd, then according to the rules of $\delta$-shift in Eq. \eref{delta-shift-1}, $S_L(z, \delta)=S_L(z)$ is independent of $\delta$, then
the corresponding propagator contributes a pole to \eref{contour-inte-delta}, localized at $z^*=-S_L/R_L$ with $S_L=\sum_{a,b\in L}s_{ab}$ and $R_L=\sum_{a,b\in L}r_{ab}$.

To evaluate such residue at $z^*=-S_L/R_L$, one can exchange the order of taking residue of $z$ and the contour integral around $\delta=\infty$ (due to the convergence of contour integral), to obtain
\bea
&&{\rm Res}_{z^*=-S_L/R_L}\,{ 1\over z(1-z^2)}\left[ {1\over 2\pi i}\,\oint_{\delta=\infty}\,{d\delta\over\delta}\, \delta^{2n-2} A_{2n}^{{\rm Tr}(\phi^3)}(\delta,z) \right]\nn
&=& {1\over 2\pi i}\,\oint_{\delta=\infty}\,{d\delta\over\delta}\, \delta^{2n-2} \left[  {\rm Res}_{z^*=-S_L/R_L}\,{ 1\over z(1-z^2)} A_{2n}^{{\rm Tr}(\phi^3)}(\delta,z) \right]\nn
&=&-\frac{1}{S_L(1-S_L^2/R_L^2)}  {1\over 2\pi i}\,\oint_{\delta=\infty}\,{d\delta\over\delta}\, \delta^{2n-2} A_{|L|+1}^{{\rm Tr}(\phi^3)}(z^*, \delta)\,A_{2n+1-|L|}^{{\rm Tr}(\phi^3)}(z^*, \delta)  \,,~~\label{step1}
\eea
where the last step uses the usual factorization of $A_{2n}^{{\rm Tr}(\phi^3)}(\delta,z)$,
\bea
\lim_{S_L(z)\to 0}\,S_L(z)\,A_{2n}^{{\rm Tr}(\phi^3)}(\delta,z)\,=\,A_{|L|+1}^{{\rm Tr}(\phi^3)}(z^*, \delta)\,A_{2n+1-|L|}^{{\rm Tr}(\phi^3)}(z^*, \delta) \,.
\eea
Suppose $|L|=2n_1-1$ and $L=\{a, a+1, \cdots, a+2n_1-2\}$ as illustrated in Fig. \ref{fig:recursion_cut}. According to the momentum conservation, we get
\begin{align}
s_{a+2n_1-2, I_L}(z^*,\delta) &= \sum_{b=a+2n_1-1}^{a-1} s_{a+2n_1-2,b}(z^*,\delta) \,,\nn
s_{I_L,a}(z^*,\delta)&=\sum_{b=a+2n_1-1}^{a-1}\,s_{b,a}(z^*,\delta)\,.
\end{align}
Hence, the momentum conservation leads to the correct $\delta$-shifts for Mandelstam variables associated with $I_L$ as
\begin{align}
s_{a+2n_1-2, I_L}(z^*,\delta) &= s_{a+2n_1-2, I_L}(z^*) -\delta, \quad s_{I_L, a}(z^*, \delta) = s_{I_L, a}(z^*) + \delta, \quad \text{if $a$ is odd,} \notag\\
s_{a+2n_1-2, I_L}(z^*,\delta) &= s_{a+2n_1-2, I_L}(z^*) +\delta, \quad s_{I_L, a}(z^*, \delta) = s_{I_L, a}(z^*) - \delta, \quad \text{if $a$ is even.}
\end{align}
In short, all Mandelstam variables of $A_{|L|+1}^{{\rm Tr}(\phi^3)}(z^*, \delta)$ have been correctly shifted in the manner of Eq. \eref{delta-shift-1}.
The case for $A_{2n+1-|L|}^{{\rm Tr}(\phi^3)}(z^*, \delta)$ is analogous.
The contour integral of $\delta$ in Eq. \eref{step1} then behaves as
\begin{align}
\label{eq:split_two_amplitude}
&{1\over 2\pi i}\,\oint_{\delta=\infty}\,{d\delta\over\delta}\, \delta^{2n-2} A_{|L|+1}^{{\rm Tr}(\phi^3)}(z^*, \delta)A_{2n+1-|L|}^{{\rm Tr}(\phi^3)}(z^*, \delta) \notag\\
=&{1\over 2\pi i}\,\oint_{\delta=\infty}\, \frac{d\delta}{\delta}  \left(\delta^{|L|-1} A_{|L|+1}^{{\rm Tr}(\phi^3)}(z^*, \delta) \right) \left(\delta^{2n-|L|-1} A_{2n+1-|L|}^{{\rm Tr}(\phi^3)}(z^*, \delta) \right) \notag\\
=& A^{\mathrm{NLSM}}_{|L|+1}(z^*) A^{\mathrm{NLSM}}_{2n+1-|L|}(z^*),
\end{align}
where the last line uses the recursive assumption: the $\delta$-shift construction is valid for NLSM amplitudes with the number of external lines no larger than $2n-2$.

Finally, we conclude the existence of
\begin{align}
{1\over 2\pi i}\,\oint_{\delta=\infty}\,{d\delta\over\delta}\, \delta^{2n-2}\,A_{2n}^{{\rm Tr}(\phi^3)}(\delta)
=& \sum_{z^*\ne0, \pm1, \infty}
A^{\mathrm{NLSM}}_{|L|+1}(z^*) \frac{1}{S_L(1-S_L^2/R_L^2)} A^{\mathrm{NLSM}}_{2n+1-|L|}(z^*)
\end{align}
which is just another form of the Eq. \eref{goal-delta}.
This completes the proof.
As can be seen, the proof is based on two crucial properties, one is that ${\rm Tr}(\phi^3)$ and NLSM amplitudes exhibit hidden zeros at the same loci in kinematic space and another is that $\delta$-shift does not affect these loci.

\subsection{Application \RNum{3}: the expansion of NLSM amplitudes}
\label{sec-expan}
In this subsection, we use the recursion relation Eq. \eref{eq:BCFW_recursion_hidden_zero} to prove that tree-level NLSM amplitudes can be expanded into bi-adjoint scalar (BAS) amplitudes with expansion coefficients being universal.
In subsection \ref{subsec-expan-bas}, we briefly review the expansion and emphasize two remarkable features of such expansion, which play crucial roles in the recursive proof. Then the details of the proof is presented in subsection \ref{subsec-proof-expan}.

\subsubsection{Expansion of NLSM amplitudes to BAS amplitudes}
\label{subsec-expan-bas}

As studied in a series of papers \cite{Feng:2019tvb,Zhou:2023quv,Zhou:2023vzl,Zhou:2024qjh}, the tree-level NLSM amplitudes can be expanded into BAS amplitudes as
\bea
A^{\rm NLSM}_{2n}(1,\cdots,2n)=\sum_{\sigma\in S_{2n-2}}\,\Big(\prod_{i\neq p,q}\,
k_i\cdot X_i\Big)\,A^{\rm BAS}_{2n}(p,\sigma_1,\cdots,\sigma_{2n-2},q|1,\cdots,2n)\,,~~\label{expan-nlsm-general}
\eea
where $p, q$ are chosen from $\{1,2,\cdots,2n\}$ and fixed at two ends of the first ordering of BAS amplitudes, $S_{2n-2}$ stands for the permutation group of the labels $\{1,\cdots,2n\}\setminus\{p,q\}$. Here $i\in \{\sigma_1,\cdots,\sigma_{2n-2}\}$ and the symbol $X_i$ is the sum of the momenta corresponding to the labels at the l.h.s of $i$ in the ordering $(p,\sigma_1,\cdots,\sigma_{2n-2},q)$. For instance, for the ordering $(1,2,3,4)$, we have $X_2=k_1$, $X_3=k_{12}\equiv k_1+k_2$.
Each tree-level BAS amplitude with two orderings consists solely of propagators for massless scalars and is simultaneously planar with respect to two orderings. The systematic method for calculating tree-level BAS amplitudes can be found in \cite{Cachazo:2013iea}.
The set of BAS amplitudes
\begin{align}
\{A^{\rm BAS}_{2n}(p,\sigma_1,\cdots,\sigma_{2n-2},q|1,\cdots,2n), \sigma\in S_{2n-2}\}
\end{align}
with $p, q\in \{1,2,\cdots,2n\}$ being fixed at two ends of the first ordering, is called the Kleiss-Kuijf (KK) basis, and the completeness of this basis is guaranteed by the well known KK relations \cite{Kleiss:1988ne}.

Let's take the simplest $4$-point amplitude $A^{\rm NLSM}_4(1,2,3,4)$ as an example to illustrate the expansion explicitly.
The expression of $4$-point amplitude $A^{\rm NLSM}_4(1,2,3,4)$ is already given in section \ref{subsec-example}. If we let $p=1$ and $q=4$, then the $4$-point BAS amplitudes in the KK basis can be evaluated as
\bea
A^{\rm BAS}_4(1,2,3,4|1,2,3,4)={1\over s_{12}}+{1\over s_{23}}\,,~~~~A^{\rm BAS}_4(1,3,2,4|1,2,3,4)=-{1\over s_{23}}\,.
\eea
Hence, we can prove that the following expansion
\bea
& &A^{\rm NLSM}_4(1,2,3,4)\nn
&=&(k_2\cdot X_2)(k_3\cdot X_3)\,A^{\rm BAS}_4(1,2,3,4|1,2,3,4)+(k_2\cdot X_2)(k_3\cdot X_3)\,A^{\rm BAS}_4(1,3,2,4|1,2,3,4)\nn
&=&(k_2\cdot k_1)(k_3\cdot k_{12})\,A^{\rm BAS}_4(1,2,3,4|1,2,3,4)+(k_2\cdot k_{13})(k_3\cdot k_1)\,A^{\rm BAS}_4(1,3,2,4|1,2,3,4)\,,~~\label{expan-4pt}
\eea
is correct. This simplest example will also serve as the starting point of recursive proof for Eq. \eref{expan-nlsm-general} in the next subsection.

To end this subsection, we highlight two significant properties of the expression on the right hand side of Eq. \eref{expan-nlsm-general}, which will be useful in the recursive proof.
\begin{itemize}
  \item Property \RNum{1}:  such expression vanishes when the total number of external legs $n$ is odd,
\begin{align}
\sum_{\sigma\in S_{n-2}}\,\Big(\prod_{i\neq p,q}\,
k_i\cdot X_i\Big)\,A^{\rm BAS}_{n}(p,\sigma_1,\cdots,\sigma_{n-2},q|1,\cdots,n) = 0.
\end{align}
  \item Property \RNum{2}: such expression is valid for any choice of $p, q\in \{1,2,\cdots,2n\}$, i.e.,
\begin{align}
&\sum_{\sigma\in S_{n-2}}\,\Big(\prod_{i\neq p,q}\,
k_i\cdot X_i\Big)\,A^{\rm BAS}_{n}(p,\sigma_1,\cdots,\sigma_{n-2},q|1,\cdots,n) \notag \\
=&\sum_{\sigma\in S_{n-2}}\,\Big(\prod_{i\neq p',q'}\,
k_i\cdot X_i\Big)\,A^{\rm BAS}_{n}(p',\sigma_1,\cdots,\sigma_{n-2},q'|1,\cdots,n).
\end{align}
\end{itemize}
As studied in \cite{Du:2018khm,Du:2019vzf}, both of two properties can be proved by using the Bern-Carrasco-Johansson (BCJ) relations for BAS amplitudes \cite{Bern:2008qj,Chiodaroli:2014xia,Johansson:2015oia,Johansson:2019dnu,Wei:2023iay}.

\subsubsection{Recursive proof}
\label{subsec-proof-expan}

The goal of this subsection is to prove the expansion in Eq. \eref{expan-nlsm-general} recursively. The expansion for $4$-point NLSM amplitude has been proved in the previous subsection, thus we assume that the expansion in Eq. \eref{expan-nlsm-general} is valid for the NLSM amplitudes with the number of external legs being no larger than $2n-2$.

Next, we need to prove Eq. \eref{expan-nlsm-general} is valid for $2n$-point NLSM amplitude.
More explicitly, applying the BCFW-like recursion in Eq. \eref{eq:BCFW_recursion_hidden_zero}, we will show
\bea
& &\sum_{\sigma\in S_{2n-2}}\,\Big(\prod_{i\neq p,q}\,
k_i\cdot X_i\Big)\,A^{\rm BAS}_{2n}(p,\sigma_1,\cdots,\sigma_{2n-2},q|1,\cdots,2n)\nn
&=&\sum_{z^*: S(z^*)=0} A_{2n_1}^{\text{NLSM}}(z^*) \frac{1}{[1-(z^*)^2]S} A_{2n+2-2n_1}^{\text{NLSM}}(z^*)\,.~~\label{goal-expan}
\eea
In the rest of this subsection, we will omit the second ordering $(1,\cdots,2n)$ of BAS amplitudes for simplicity, if it doesn't result in ambiguity.
We apply the $z$-shift and contour integration used to prove Eq. \eref{eq:BCFW_recursion_hidden_zero} to the first line of \eref{goal-expan} and obtain
\bea
& &\sum_{\sigma\in S_{2n-2}}\,\Big(\prod_{i\neq p,q}\,
k_i\cdot X_i\Big)\,A^{\rm BAS}_{2n}(p,\sigma_1,\cdots,\sigma_{2n-2},q)\nn
&=&{1\over 2\pi i}\,\oint_{|z|=\epsilon}\,{dz\over z}\,{1\over 1-z^2}\,\left[\sum_{\sigma}\,\Big(\prod_{i\neq p,q}\,
T_i(z)\Big)\,A^{\rm BAS}_{2n}(p,\sigma_1,\cdots,\sigma_{2n-2},q;z)\right]\nn
&=&-\sum_{z^*\ne0}\,{\rm Res}_{z=z^*}\,{1\over z(1-z^2)}\,\left[\sum_{\sigma}\,\Big(\prod_{i\neq p,q}\,
T_i(z)\Big)\,A^{\rm BAS}_{2n}(p,\sigma_1,\cdots,\sigma_{2n-2},q;z)\right]\,,~~\label{sum-res-expan}
\eea
where $T_i\equiv k_i\cdot X_i$.

Being similar with subsection \ref{subsec-prove-delta}, we find that the residues at poles $z=\infty,\pm1$ vanish.

\textbf{Residue for pole $z=\infty$~} For the pole at $z=\infty$, $\left(\prod_{i\ne p,q} T_i(z) \right) A^{\rm BAS}_{2n}(z)$ has mass dimension $2$ and is proportional to $z$ in the large $z$ limit. Hence, following a similar reason is subsection \ref{subsec-prove-delta}, we find the residue is $0$.

\textbf{Residue for pole $z=\pm1$~}
For the poles at $z=\pm1$, we show that the expansion in the first line of Eq. \eref{goal-expan} exhibits the same hidden zeros as the NLSM amplitude, namely
\bea
\sum_{\sigma\in S_{2n-2}}\,\Big(\prod_{i\neq p,q}\,
T_i\Big)\,A^{\rm BAS}_{2n}(p,\sigma_1,\cdots,\sigma_{2n-2},q)\,\xrightarrow[]{\eref{kinematic-condi-zero}}\,0\,.~~\label{zero-for-expan}
\eea
According to the property \RNum{2} mentioned at the end of the previous subsection, we can let $\{p,q\}=\{\tilde{i},\tilde{j}\}$, where $\tilde{i}$ and $\tilde{j}$ are two reference legs used to define the hidden zero kinematics Eq. \eref{kinematic-condi-zero}, and split the remaining external legs into two sets $A$ and $B$ according to the ordering $(1,2,\cdots,2n)$. Based on this choice, we can reorganize the first line of Eq. \eref{goal-expan} as
\bea
& &\sum_{\sigma\in S_{2n-2}}\,\Big(\prod_{i\neq \tilde{i},\tilde{j}}\,
T_i\Big)\,A^{\rm BAS}_{2n}(\tilde{i},\sigma_1,\cdots,\sigma_{2n-2},\tilde{j})\nn
&=&\sum_{\alpha_A,\beta_B}\,\sum_{\shuffle}\,\Big(\prod_{a\in A}\,
T_a\Big)\,\Big(\prod_{b\in B}\,
T_b\Big)\,A^{\rm BAS}_{2n}(\tilde{i},\alpha_A\shuffle\beta_B,\tilde{j}).~~\label{zero-expan}
\eea
Here $\alpha_A$ and $\beta_B$ are the permutations of $A$ and $B$ respectively and the first summation is over all possible $\alpha_A$ and $\beta_B$, and the summation for the shuffle $\shuffle$ is taken over all permutations of $\alpha_A\cup\beta_B$ such that the orderings of $\alpha_A$ and $\beta_B$ are kept. For instance, suppose $\alpha_A=\{1,2\}$, $\beta_B=\{3,4\}$, then
the summation of $\shuffle$ is
\bea
\sum_{\shuffle}\,A(\tilde{i},\alpha_A\shuffle\beta_B,\tilde{j})&=&A(\tilde{i},1,2,3,4,\tilde{j})+A(\tilde{i},1,3,2,4,\tilde{j})+A(\tilde{i},1,3,4,2,\tilde{j})\nn
& &+A(\tilde{i},3,4,1,2,\tilde{j})+A(\tilde{i},3,1,4,2,\tilde{j})+A(\tilde{i},3,1,2,4,\tilde{j})\,.
\eea
Here we also split $\{T_i\}$ into two subsets, $\{T_a\}$ and $\{T_b\}$, according to $A$ and $B$.
According to the kinematics of the hidden zero of $(\tilde{i}, \tilde{j})$, $k_a\cdot k_b=0$ for $a\in X_a$ and $b\in B$, thus $T_a$ is independent of $\beta_B$ and $T_b$ is independent of $\alpha_A$.
Hence, the summation of the shuffle $\shuffle$ can be moved to the r.h.s of the factor of $T_a$ and $T_b$, as following
\begin{align}
&\sum_{\alpha_A,\beta_B}\,\sum_{\shuffle}\,\Big(\prod_{a\in A}\,
T_a\Big)\,\Big(\prod_{b\in B}\,
T_b\Big)\,A^{\rm BAS}_{2n}(\tilde{i},\alpha_A\shuffle\beta_B,\tilde{j}) \notag\\
\overset{{\eref{kinematic-condi-zero}}}{=} &\sum_{\alpha_A,\beta_B}\,\Big(\prod_{a\in A}\,
T_a\Big)\,\Big(\prod_{b\in B}\,
T_b\Big)\,\Big[\sum_{\shuffle}\,A^{\rm BAS}_{2n}(\tilde{i},\alpha_A\shuffle\beta_B,\tilde{j})\Big].
\end{align}
Further, we can apply the KK relations in \cite{Kleiss:1988ne} to the BAS amplitudes as
\bea
\sum_{\shuffle}\,A^{\rm BAS}_{2n}(\tilde{i},\alpha_A\shuffle\beta_B,\tilde{j})=(-1)^{|B|}\,A^{\rm BAS}_{2n}(\tilde{i},\alpha_A,\tilde{j},\beta_B^T)\,,~~\label{kk-relation}
\eea
where $|B|$ denotes the cardinality of the set $B$ and $\beta_B^T$ stands for the reverse of $\beta_B$ like, $\beta_B^T=\{3,2,1\}$ for $\beta_B=\{1,2,3\}$.
Since two subsets $A$ and $B$ are separated by $\tilde{i}$ and $\tilde{j}$ in both orderings $(\tilde{i},\alpha_A,\tilde{j},\beta_B^T)$ and $(1,2,\cdots,2n)$, thus the BAS amplitude $A^{\rm BAS}_{2n}(\tilde{i},\alpha_A,\tilde{j},\beta_B^T|1,\cdots,2n)$ at the r.h.s of Eq. \eref{kk-relation} vanishes on the locus Eq. \eref{kinematic-condi-zero} \cite{Zhang:2024efe, Huang:2025blb}. So we successfully prove Eq. \eref{zero-for-expan}.

Since the kinematics of the shifted Mandelstam variables at poles $z=\pm1$ correspond to two hidden zeros in Eq. \eref{realize-zero}, thus we conclude that the residues at poles $z=\pm1$ vanish.

\textbf{Residues for other physical poles~}
Now the expansion can be calculated by summing the residues at physical poles as
\bea
& &\sum_{\sigma\in S_{2n-2}}\,\Big(\prod_{i\neq p,q}\,
T_i\Big)\,A^{\rm BAS}_{2n}(p,\sigma_1,\cdots,\sigma_{2n-2},q)\nn
&=&-\sum_{z^*:S(z^*)=0}\,{\rm Res}_{z=z^*}\,{1\over z(1-z^2)}\,\left[\sum_{\sigma}\,\Big(\prod_{i\neq p,q}\,
T_i(z)\Big)\,A^{\rm BAS}_{2n}(p,\sigma_1,\cdots,\sigma_{2n-2},q;z)\right]\,,~~\label{sum-res-expan-2}
\eea
where the poles at $z=z^*$ are physical poles of BAS amplitudes $A^{\rm BAS}_{2n}(p,\sigma_1,\cdots,\sigma_{2n-2},q;z)$ and $S(z)$ is the shifted planar Mandelstam variable.
Recall that each BAS amplitude corresponds to two ordering $(p,\sigma_1,\cdots,\sigma_{2n-2},q)$ and $(1,\cdots,2n)$ and each Feynman diagram contributing to this amplitude should be simultaneously planar with respect to both orderings, thus the planar Mandelstam variable $S(z)$ contributing to the amplitude should has the form $s_{f\cdots g}(z)$, which corresponds to a set of consecutive external legs $\{f,f+1,\cdots,g-1,g\}$. For a pair of external legs $\{f,g\}$ corresponding to a planar Mandelstam variable $s_{f\cdots g}(z)$, we can use the the property \RNum{2} discussed at the end of previous subsection to rewrite the expansion as
\bea
& &\sum_{\sigma}\,\Big(\prod_{i\neq p,q}\,
T_i(z)\Big)\,A^{\rm BAS}_{2n}(p,\sigma_1,\cdots,\sigma_{2n-2},q|1,2,\cdots,2n;z)\nn
&=&\sum_{\sigma}\,\Big(\prod_{i\neq f-1,f}\,
T_i(z)\Big)\,A^{\rm BAS}_{2n}(f,\sigma_1,\cdots,\sigma_{2n-2},f-1|f,\cdots,f-1;z)\,.~~\label{BAS-fg}
\eea
where we have used the cyclic symmetry of the second ordering $(1,2,\cdots,2n)$.
Then $g$ splits the ordering $(\sigma_1,\cdots,\sigma_{2n-2})$ into two sub-orderings $\gamma_1=(a_1,\cdots,a_{|L|})$ and $\gamma_2=(b_1,\cdots,b_{|R|})$, where $a_i\in\{f+1,\cdots,g\}$ and $b_j\in\{g+1,\cdots,f\}$. Each BAS amplitude containing the propagator $s_{f\cdots g}(z)$ factorizes as
\begin{align}
s_{f\cdots g}(z)&A^{\rm BAS}_{2n}(f,\gamma_1,\gamma_2,f-1|f,\cdots,f-1;z)\,\xrightarrow[]{s_{f\cdots g}(z)=0}\nn
&A^{\rm BAS}_{n'_1}(f,\gamma_1,I_L|f,\cdots,g,I_L;z)\,A^{\rm BAS}_{2n+2-n'_1}(I_R,\gamma_2,f-1|I_R,g+1,\cdots,f-1;z)\,,
\end{align}
where $I_R, I_L$ are two legs produced by cutting the internal line $s_{f\cdots g}(z)$.
At the same time, the factor $T_i(z)$'s can also be classified into two types, $T_\ell$ and $T_r$ with $\ell\in \gamma_1$ and $r\in \gamma_2$. For instance, for $A^{\rm BAS}_{2n+2-n'_1}(I_R,1,2,f-1;z^*)$, $T_r(z)$ are represented as
\bea
T_1(z)={1\over2}\,s_{1I_R}(z)\,,\quad T_2(z)={1\over2}\,s_{2I_R}(z)+{1\over2}\,s_{12}(z).
\eea
Therefore, we can compute the residues in Eq. \eref{sum-res-expan-2} and obtain
\bea
& &\sum_{\sigma\in S_{2n-2}}\,\Big(\prod_{i\neq p,q}\,
T_i\Big)\,A^{\rm BAS}_{2n}(p,\sigma_1,\cdots,\sigma_{2n-2},q)\nn
&=&\sum_{z^*:s_{f\cdots g}(z^*)=0}\,{1\over 1-(z^*)^2}\,\left[\sum_{\gamma_1}\,\Big(\prod_{\ell\in\{f+1,\cdots,g\}}\,T_\ell(z^*)\Big)\,A^{\rm BAS}_{2n_1}(f,\gamma_1,I_L|f,\cdots,g,I_L;z^*)\right]\nn
& &{1\over s_{f\cdots g}}\,\left[\sum_{\gamma_2}\,\Big(\prod_{r\in\{g+1,\cdots,f-2\}}\,
T_r(z^*)\Big)\,A^{\rm BAS}_{2n+2-2n_1}(I_R,\gamma_2,f-1|I_R,g+1,\cdots,f-1;z^*)\right]\,,~~\label{sum-res-expan-3}
\eea
where the terms corresponding to $n'_1$ being odd vanish, because of the property \RNum{1} mentioned at the end of previous subsection.
Because of the recursive assumption,
\begin{align}
A_{2n_1}^{\mathrm{NLSM}}(f,\cdots,g,I_L;z^*) &= \sum_{\gamma_1}\,\Big(\prod_{\ell\in\{f+1,\cdots,g\}}\,T_\ell(z^*)\Big)\,A^{\rm BAS}_{2n_1}(f,\gamma_1,I_L|f,\cdots,g,I_L;z^*)   \notag\\
A_{2n_1}^{\mathrm{NLSM}}(I_R,g+1,\cdots,f-1;z^*) &= \sum_{\gamma_2}\,\Big(\prod_{r\in\{g+1,\cdots,f-2\}}\,
 T_r(z^*)\Big)\, A^{\rm BAS}_{2n+2-2n_1}(I_R,\gamma_2,f-1|I_R,g+1,\cdots,f-1;z^*) \notag
\end{align}
we finally obtain
\begin{align}
&\sum_{\sigma\in S_{2n-2}}\,\Big(\prod_{i\neq p,q}\,
T_i\Big)\,A^{\rm BAS}_{2n}(p,\sigma_1,\cdots,\sigma_{2n-2},q)\nn
=&\sum_{z^*:s_{f\cdots g}(z^*)=0} A_{2n_1}^{\mathrm{NLSM}}(f,\cdots,g,I_L;z^*)
\frac{1}{s_{f\cdots g}[1-(z^*)^2]} A_{2n_1}^{\mathrm{NLSM}}(I_R,g+1,\cdots,f-1;z^*).
\end{align}
So we complete the proof of Eq. \eref{goal-expan} successfully.

\section{Conclusion and discussion}
\label{sec-conclu}

In this note, we propose a novel recursion relation for the tree-level NLSM amplitudes, which avoids boundary terms by exploiting the hidden zeros. Just as the ordinary BCFW recursion relations, the new recursion relation Eq. \eref{eq:BCFW_recursion_hidden_zero} is involved with only the factorization on physical poles, while the recursion relation in \cite{Cachazo_2022} needs the off-shell amputated currents. Such concise recursion relation allows us to prove several interesting properties of tree-level NLSM amplitudes: the Adler zero, the $\delta$-shift construction, and the universal expansion into BAS amplitudes.

An important problem in the modern S-matrix program is to find the minimal set of physical fundamental principles which can uniquely determine the scattering amplitudes of some theories. in this note, the BCFW-like recursion relation Eq. \eref{eq:BCFW_recursion_hidden_zero} clearly shows that the hidden zero, together with the factorization on physical poles, completely determine the NLSM amplitudes.
In fact, the $\delta$-shift construction and the expansion into BAS amplitudes are also proved via these two principles, since the proof for each case depends on two key steps, the first is to show the absence of poles at $z=\pm1,\infty$ by using the hidden zero and the second is to compute the residues at finite physical poles.


The BCFW-like recursion relation in Eq. \eref{eq:BCFW_recursion_hidden_zero} may also be applied to establish other properties of NLSM amplitudes, such as the Bern-Carrasco-Johansson (BCJ) relations \cite{Chen:2013fya} and the recently discovered $2$-split factorization behavior \cite{Cao:2024gln,Cao:2024qpp}. Owing to the substantial technical challenges involved, we defer these investigations to future work.


Another interesting future direction is to extend the recursion relation proposed in this paper to Feynman integrands at the loop level. As can be anticipated, a crucial ingredient in this extension is the hidden zeros at the loop level, since they play a key role in eliminating boundary terms. The hidden zeros of $1$-loop NLSM Feynman integrands have been recently investigated in \cite{Backus:2025hpn}, which employs surface kinematics. This work provides a solid foundation for generalizing our recursion relation to the $1$-loop level.

\section*{Acknowledgments}

We thank Prof. Bo Feng and Prof. Yijian Du for extremely useful discussions and suggestions. KZ is supported by NSFC under Grant No. 11805163.

\appendix

\section{Proof for Eq. \eref{expan-s}}
\label{sec-Appen}

To support the expansion Eq. \eref{expan-s} used in section \ref{subsec-prove-delta}, we now give a proof of this formula.
On can always sum all terms of $A_{2n}^{{\rm Tr}(\phi^3)}$ together to get a rational function
\bea
A_{2n}^{{\rm Tr}(\phi^3)}={P\over M}\,,
\eea
where the numerator $P$ is a polynomial of Mandelstam variables, and the denominator $M$ is a monomial consists of physical poles.
Suppose the kinematics of a hidden zero $\{\tilde{i}, \tilde{j}\}$ contains $\{s_{ab}=0, a\in A, b\in B\}$ as in Eq. \eref{kinematic-condi-zero}. Since $a$ and $b$ are not consecutive, the amplitude will not contain such propagator $1/s_{ab}$, thus the denominator $M$ can never be expressed as $M=s_{ab}M'$ for any $a\in A$ and $b\in B$, where $M'$ is also a monomial. In other words, $M\ne 0$ in the kinematics of hidden zero $\{\tilde{i}, \tilde{j}\}$. Hence, the hidden zero of $A_{2n}^{{\rm Tr}(\phi^3)}$ indicates that $P=0$ in the kinematics of hidden zero $\{\tilde{i}, \tilde{j}\}$.

For simplicity, let us denote these $s_{ab}$'s as $s_i$ with $i\in\{1,\cdots,h\}$ with $h$ being the number of $s_{ab}$'s. Further more, we express $P$ as the polynomial of $n(n-3)/2$ independent Mandelstam variables which involve $\{s_1,\cdots,s_h\}$ as a subset, to avoid the subtleties caused by momentum conservation. Then, the polynomial $P$ can be decomposed as
\bea
P=s_1\,D_1+P_{2\cdots h}\,,
\eea
where the first term $s_1D_1$ is obtained by collecting all monomials containing $s_1$ together (if $P$ is independent of $s_1$, then $D_1=0$). After performing the above operation, the polynomial $P_1$ is clearly independent of $s_1$. To proceed, we split $P_1$
as $P_{2\cdots h}=s_2D_2+{P_{3\cdots h}}$ following the same manner, which yields
\bea
P=s_1\,D_1+s_2\,D_2+P_{3\cdots h}\,.
\eea
Repeating the above manipulations, we can arrive at
\bea
P=\sum_{i=1}^{h}\,s_i\,D_i+P_0\,,~~\label{P-expan}
\eea
where $P_0$ does not contain any $s_i$ with $i\in\{1,\cdots,h\}$.
Because the hidden zero indicates $P=0$, then $P_0$ must also vanish in the kinematics of hidden zero $\{\tilde{i}, \tilde{j}\}$. Meanwhile, since $P_0$ is independent of any $s_i$, and the freedom of reformulating based on momentum conservation was removed at the beginning. Thus the changing of $s_i$ will not affect $P_0$, and we find $P_0=0$. So we conclude that $P=\sum_{i=1}^{h}\,s_i\,D_i$, which is equivalent to \eref{expan-s} if we set $B_i=D_i/M$.

\bibliographystyle{JHEP}
\bibliography{ref}

\end{document}